\documentclass[preprint2]{aastex}

\usepackage{graphicx}		
\usepackage{color}			
\usepackage{amsmath}
\usepackage[breaklinks=true]{hyperref}
\usepackage{breakcites}

\shorttitle{Explosive SNIa and GT Strengths}
\shortauthors{Mori et al.}

\begin{document}
\title{
	Impact of New Gamow-Teller Strengths on Explosive Type Ia 
	Supernova Nucleosynthesis$^*$
	}
	
\author{Kanji Mori\altaffilmark{1,2}, Michael A. Famiano\altaffilmark{1,3}, Toshitaka Kajino\altaffilmark{1,2},
	Toshio Suzuki\altaffilmark{1,4}, Jun Hidaka\altaffilmark{5}, Michio Honma\altaffilmark{6}, 
	Koichi Iwamoto\altaffilmark{7}, Ken'ichi Nomoto\altaffilmark{8,9}, Takaharu Otsuka\altaffilmark{10,11}}
\email{kanji.mori@nao.ac.jp}
\email{michael.famiano@wmich.edu} 
\email{kajino@nao.ac.jp}
\email{suzuki@phys.chs.nihon-u.ac.jp}
\email{jun.hidaka@meisei-u.ac.jp}
\email{m-honma@u-aizu.ac.jp}
\email{iwamoto@phys.cst.nihon-u.ac.jp}
\email{nomoto@astron.s.u-tokyo.ac.jp}
\email{otsuka@phys.s.u-tokyo.ac.jp}
	\altaffiltext{1}{
		National Astronomical Observatory of Japan 2-21-1 
		Osawa, Mitaka, Tokyo, 181-8588 Japan}
	\altaffiltext{2}{Department of Astronomy, Graduate School of Science, 
		The University of Tokyo, 7-3-1 Hongo, Bunkyo-ku, Tokyo, 113-0033 Japan}
	\altaffiltext{3}{Department of Physics, Western Michigan University, Kalamazoo, Michigan 49008 USA}
	\altaffiltext{4}{
		Department of Physics, College of Humanities and Sciences,
		Nihon University 3-25-40 Sakurajosui, Setagaya-ku, Tokyo 156-8550 Japan}
	\altaffiltext{5}{Mechanical Engineering Department,
		Mesei University, 2-1-1 Hodokubo, Hino, Tokyo 191-8506 Japan}
	\altaffiltext{6}{Center for Mathematical Sciences, University of Aizu, Aizu-Wakamatsu, Fukushima 965-8580 Japan}
	\altaffiltext{7}{Department of Physics , College of Science and Technology, Nihon University,
		Tokyo 101-8308 Japan}
	\altaffiltext{8}{Kavli Institute for the Physics and Mathematics of the Universe (WPI), The University of Tokyo, Kashiwa, Chiba 277-8583 Japan}
	\altaffiltext{9}{Hamamatsu Professor}	
	\altaffiltext{10}{
		Department of Physics, Graduate School of Science, 
		The University of Tokyo, 7-3-1 Hongo, Bunkyo-ku, Tokyo, 113-0033 Japan}
	\altaffiltext{11}{National Superconducting Cyclotron Laboratory, Michigan State University, East Lansing, MI 48824 USA
	}
	\altaffiltext{*}{Accepted to Astrophysical Journal 17-October-2016}

\begin{abstract}
Recent experimental results have confirmed a possible reduction in the GT$_+$
strengths of pf-shell nuclei.  These proton-rich nuclei are of relevance in the
deflagration and explosive burning phases of Type Ia supernovae.  While prior
GT strengths result in nucleosynthesis predictions with a lower-than-expected
electron fraction, a reduction in the GT$_+$ strength can result in an
slightly increased
electron fraction compared to previous shell model predictions, though
	the enhancement is not as large as previous enhancements in going from
	rates computed by Fuller, Fowler, and Newman based on an independent 
	particle model.  A shell model parametrization has been developed which more
closely matches experimental GT strengths.  The resultant electron-capture rates
are used in nucleosynthesis calculations for carbon deflagration and explosion phases
of Type Ia supernovae, and the final mass fractions are compared to those 
obtained using more commonly-used rates.
\end{abstract}


\maketitle 

\section{Introduction}
\label{intro}
Type Ia supernovae are thought to result from accreting C-O white dwarfs (WDs) in close binaries \citep[e.g.,][]{hoyle60,arnett96,hillebrandt00,boyd08,illiadis12}. (Here we denote type Ia supernovae as ``SNe Ia'' and a single
type Ia supernova as ``SN Ia.'')
If the WD reaches a certain critical condition, thermonuclear burning ignited in the electron-degenerate matter results in a cataclysmic explosion of the whole star.  
Material that is abundant in Fe-peak elements, including some neutron-rich ones, is ejected into the interstellar medium (ISM), contributing to chemical enrichment in galaxies. SNe Ia also play an important role in cosmology to measure the expansion rate of the Universe \citep{riess98,perlmutter99,schmidt08}.
\subsection{Type Ia Supernovae}
The formation of SNe Ia and their progenitors have been an issue of debate
\citep[e.g.,][]{maoz14,hillebrandt00,nomoto95}.
In typical cases of accretion from a non-degenerate companion star, known as the single-degenerate (SD) model, the WD mass approaches the
Chandrasekhar mass limit and SNe Ia are induced (``Chandra model'').
In double-degenerate (DD) models, two WDs merge to produce a SN Ia 
\citep{iben84,webbink84}; in recent violent merger models \citep[e.g.,][]{pakmor,sato16}, thermonuclear explosions are induced in sub-Chandrasekhar-mass WDs (``sub-Chandra model'').
Important differences between the two models are 
the central densities ($\rho_c$) of the exploding WDs.  In the Chandra model,
$\rho_c > 10^9$ g cm$^{-3}$, while $\rho_c \lesssim 10^8$ g cm$^{-3}$
in the sub-Chandra models \citep{wang12,nomoto11}.   

In Chandra models, thermonuclear burning triggered in the central region of the WD propagates outward as a subsonic deflagration (flame) 
front \citep{nomoto76,nomoto84}. Rayleigh-Taylor instabilities 
at the flame front cause the development of turbulent eddies, which increase the flame surface area,  enhancing the net burning rate and accelerating the flame 
\citep{muller82,arnett94,khokhlov95}.  In some cases the deflagration may be strong enough to undergo a deflagration to
detonation transition \citep[DDT:][]{blinnikov86,khokhlov91,nomoto}.
The turbulent nature of the flame propagation including the possible DDT
and associated nucleosynthesis have been studied in full 3D simulations
\citep[e.g.,][]{gamezo05,roepke06,seitenzahl13}.

The main observable characteristics of SNe Ia are their optical light curves and spectra. 
The light curves are powered primarily 
via the decays of \(^{56}\)Ni and its daughter \(^{56}\)Co \citep{arnett79}.  
The early spectra are characterized by the presence 
of strong absorption lines of Si; the intermediate-mass elements such as Ca, S, Mg, and O; and 
the Fe-peak elements Fe, Ni, and Co 
\citep{branch85,parrent14}.
 The late-time spectra show emission lines of Fe-peak elements, which include those of stable
 Ni, i.e., neutron-rich $^{58}$Ni.
It is thus evident that the light curves and spectra are closely related to the nucleosynthesis, 
which is crucial to study the unresolved issues regarding the explosion models and the progenitors of SNe Ia. 

Explosive nucleosynthesis calculations in both the Chandra and sub-Chandra models predict the
production of reasonable amounts of Fe-peak elements and intermediate mass
elements (Ca, S, Si, Mg, O).  In the inner parts of the WD,
temperatures behind the deflagration or detonation wave exceed \(\sim
5 \times 10^9\) K, so that the reactions are rapid enough to
synthesize mainly \(^{56}\)Ni.  In the surrounding parts
with lower densities and temperatures, explosive burning produces the
intermediate mass elements Si, S, Ar, Ca.  Both models are
successful in producing the basic features of SN Ia light
curves and spectra.  This implies that it is difficult to distinguish
these models.

However, the synthesized amounts of some neutron-rich species, such as
\(^{58}\)Ni, \(^{54}\)Fe, and \(^{55}\)Mn relative to \(^{56}\)Fe
differ between the Chandra and sub-Chandra models because of the
difference in the central densities of the WDs \citep{yamaguchi15}.
\subsection{Nuclear Physics Inputs to SNe Ia Models}
In the Chandra models, the central densities of the WDs are high enough that 
the Fermi energy of electrons tends to 
exceed the energy thresholds of the electron captures involved. 
Electron captures reduce the electron mole fraction, \(Y_e\), 
that is the number of electrons per baryon,
\begin{equation}
Y_e \equiv \sum_i Z_i Y_i,
\end{equation}
where the sum is over all nuclear species, and \(Y_i\) is the abundance for a given 
species \(i\) with \(Z_i\) protons. As a result of electron capture, the Chandra-model synthesizes a significant
amount of neutron-rich Fe-peak elements.  The detailed
abundance ratios with respect to \(^{56}\)Ni (or \(^{56}\)Fe) depend
on the convective flame speed and the central densities 
(e.g.,
\citet{benvenuto15} for rotating WD models), 
which
must be studied in multi-D hydrodynamical simulations.

On the other hand, the sub-Chandra models undergo little electron capture,
so that the amount of stable Ni is much smaller.
Such a difference in densities at nuclear burning can be tested with 
various observations. Specifically, the following observations are sensitive to the central density of the WD, and hence, whether the model is a Chandra or sub-Chandra model.
\begin{enumerate}
\item The late time spectra of some SNe Ia show features of stable Ni and Fe \citep[e.g.,][]{maede10c,nomoto11}.
\item X-ray spectra of SN Ia remnants provide abundance ratios such
as stable Ni and Mn with respect to Fe \citep{yamaguchi15}.
\item Solar abundance patterns of Fe-peak isotopes would constrain the
ratios of Ni/Fe, Mn/Fe, etc. \citep[e.g.,][]{nomoto11}, depending on the chemical evolution
models of our Galaxy and the produced abundances in indiviual type Ia events \citep[e.g.,][]{nomoto15,seitenzahl13}.
\end{enumerate}

To constrain the explosion conditions and thus the explosion models,
it is important to accurately predict the electron capture rates relevant to nucleosynthesis in SNe Ia.

Most of the nucleosynthesis studies so far employ electron capture
rates based on shell model estimations when available
\citep{ffn1,langanke01}, most of which are simpler than what is currently
possible. Nuclear physics experiments and improved
theoretical descriptions of weak rates \citep[for example,][]{kbf}
have constrained SN Ia nucleosynthesis in attempts to accurately predict
the abundance ratios of Fe-peak nuclei. These "second
generation" results have greatly improved upon earlier
nucleosynthesis predictions \citep{martinez03}.

The current push in describing SN Ia nucleosynthesis is towards
reducing the uncertainty in the yields while also attempting to
replicate experimental results which are now made possible by improved
techniques \citep[for example,][]{sasano11,honma04}. Work in
this direction will provide not only convergence in nuclear models
towards an accurate description of weak rates, but also a measure of
the sensitivity of SN Ia models to the nuclear physics inputs.

Because of their relevance to SNe Ia, the
	GT strengths of pf-shell nuclei have undergone significant
	experimentatal investigations over the past several years.  
	Various techniques have been employed to extract the
	GT strengths in the pf-shell nuclei.  A compilation of 
	GT measurements of many of the pf-shell nuclei
	with 45$\le Z \le 64$ was performed by \cite{cole12}. This study includes 
	studies of $\beta$-decay results \citep{alford91, williams95,
		anantaraman08, popescu07},
	charge exchange results using (n,p) reactions
	\citep{yako09, vetterli87, vetterli89, anantaraman08}, charge-exchange results using
	(d,$^2$He) reactions
	\citep{hitt09, baumer05, alford93, baumer03,cole06},
	 charge-exchange results using
	 (t,$^3$He) reactions
	 \citep{hagermann04, hagermann05, 
	  grewe08},
	 and charge-exchange results using (p,n) reactions assuming isospin symmetry
	 \citep{williams95, anantaraman08}.

In addition to experimental work, there has been theoretical interest in computing GT strengths.  In nuclear reaction networks, the commonly-used KBF model \citep{kbf} is often employed to generate EC rates.
	The KB3G and KBF rates are generally
	the standards used in nucleosynthesis calculations \citep{langanke01} and
	have been for over a decade.  
	However, more recent models have been proposed, including
	the GXP-type shell model family \citep{suzuki11}.  These two models 
	have been found to produce GT strength distributions which are in disagreement, resulting in potentially different
	EC rates, and thus possible different nucleosynthetic yields in SNe Ia. Which model is correct is the subject of the aforementioned experimentation.

As an example, recent measurements of the GT transition strength 
distributions for the $^{56}$Ni$\rightarrow ^{56}$Cu and the
$^{55}$Co$\rightarrow ^{55}$Ni transitions \citep{sasano12,sasano11} are helpful in 
constraining nuclear shell models. Using the
(p,n) charge-exchange reaction, the strengths were measured
with a good degree of accuracy over a very wide range of excitation
energies.  Here, isospin symmetry-breaking effects are 
neglected, and the experimentally extracted GT strengths
for $^{56}$Ni$\rightarrow  ^{56}$Cu can be applied to electron
capture (EC) and $\beta^+$ decays of $^{56}$Ni \citep{sasano11}. It was found that the transition strengths for these
two transitions more closely match those computed with the
GXPF1J shell-model interaction \citep{suzuki11,honma04} than those computed
with the KB3G or the KBF models \citep{kb3g,kbf}. 

These experimental and theoretical results suggest that actual EC rates may differ from those predicted by the KBF-family shell model calculations.
	A reduction of the EC rates
may be responsible for an overall 
enhancement  in $Y_e$ and a concurrent change in production of $^{56}$Ni in SNe Ia.  The effect may be magnified if 
all nuclei in the pf-shell region are considered.  Here, strength
calculations result in EC rates that are lower for many pf-shell nuclei (for $\rho Y_e=10^7$ g~cm$^{-3}$) \citep{suzuki11}.  
However, depending on the temperature and $\rho Y_e$ of the medium, rates may increase or decrease as
	different parts of the GT strength function are integrated over. While decreases in EC rates may change the
	production of $^{56}$Ni in SNe Ia, increased rates cannot be ruled out from the data. This was verified by \cite{cole12}, who found no systematic shift in
	rates if one shell model is used over another. Depending on the shell
	model, strengths may be higher or lower for different
	thermal conditions.  This concept will be discussed later in
	this paper.

The effects of the GXP-type shell model on proton-rich pf-shell nuclei with 23$\le$Z$\le$30 are studied as they influence nucleosynthesis 
in SNe Ia.  Here, we examine both stable and unstable nuclei in this region.  In particular, the effect on $Y_e$ as well as production
ratios are evaluated.  Mass fractions of nuclei produced in SNe Ia are compared using both GXP parametrizations and KBF models. Trajectories
of mass shells in a WD are used as input into a nuclear reaction
network to gauge the effects of variations in nuclear physics inputs, and final nuclear mass fractions in individual shells are computed.
Because of computational limitations, the explosion calculation is decoupled from the nuclear reaction network.  However, the effects
of the nuclear shell model used are evident in the resultant electron
fractions and the final mass fractions.  Comparisons to solar
values indicate that the enhancement in $Y_e$, which arises from
using the GXP-type model, reduces the overall $^{58}$Ni/$^{56}$Ni and
$^{58}$Ni/$^{56}$Fe ratios, which has been an interesting problem 
addressed by prior evaluations \citep{brachwitz00,nomoto}.

A brief overview of the GXP shell model calculation is
 presented in \S\ref{shell_calcs} along with a
  comparison to KBF rates.  Next, the nuclear reaction network
    calculations
and the insertion of the GXP rates are presented.  Results, including
produced mass fractions are compared in \S\ref{results}; these
results using the GXP shell model are compared to results using
the KBF rates as well as solar mass fractions. Finally, conclusions,
discussion, and future prospects are presented.
\section{Electron Capture Rate Calculations} 
\label{shell_calcs}
The GXPF1J Hamiltonian \citep{honma05} was modified from the original
GXPF1 Hamiltonian \citep{honma04}, which was obtained by fitting experimental
energy data of a wide range of pf-shell nuclei with mass number 
$47\le A\le 66$. New experimental data of neutron-rich Ca, Ti and Cr 
isotopes with $N >$ 32 were taken into account to improve the GXPF1J model.
This model was further modified to reproduce the peak position
of the M1 strength in $^{48}$Ca.  The KBF and KB3G give energies   
for the 1$^{+}$ state about 1 MeV below the experimental value.
In the KB3G model the M1 strength is split into two states.
The M1 strengths in $^{50}$Ti, $^{52}$Cr and $^{54}$Fe as well as the
GT strength in $^{58}$Ni are found to be well reproduced by GXPF1J 
with the use of the quenching factors, $g_{s}^{eff}$/$g_{s}$ 
=0.75 \citep{cosel98} and $g_{A}^{eff}$/$g_{A}$ =0.74 \citep{caurier05}, 
respectively.
\begin{figure*}
	\includegraphics[width=0.49\linewidth]{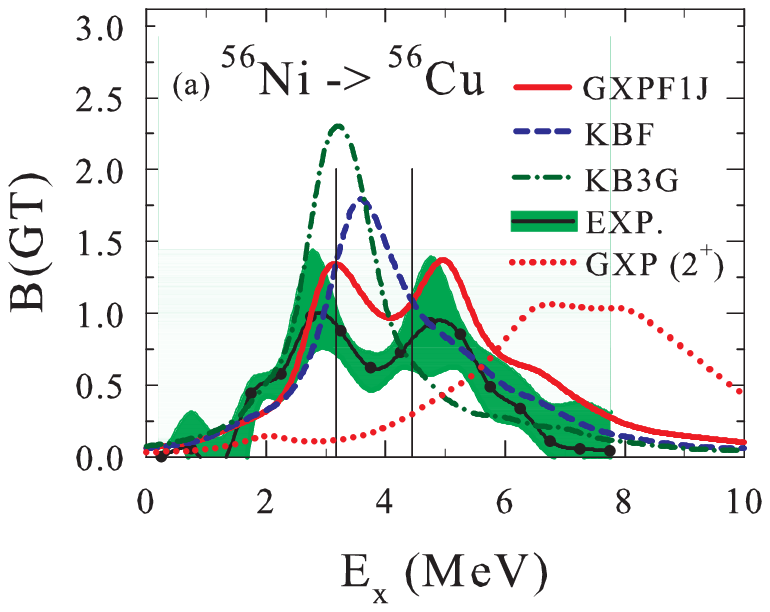}
	\includegraphics[width=0.49\linewidth]{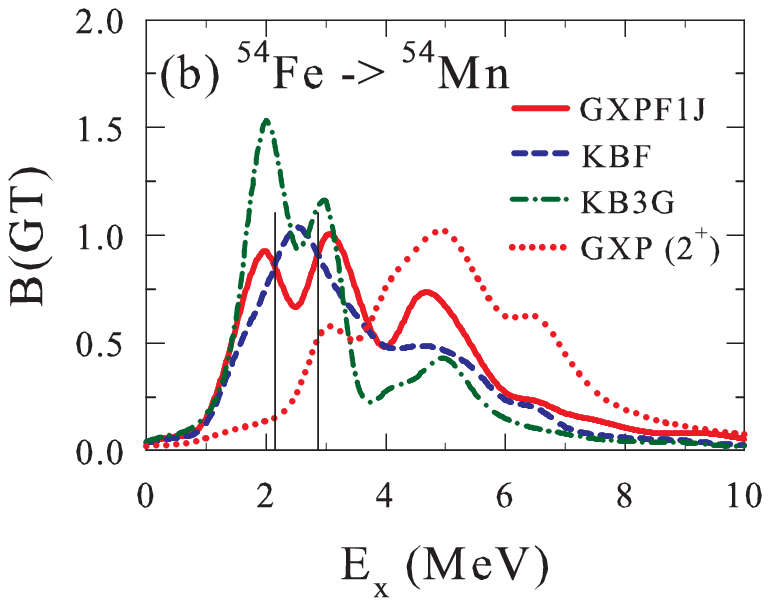}
	\caption{Comparison of the B(GT) strength function. (a) Three shell
		model calculations and experimental data for $^{56}$Ni$\rightarrow^{56}$Cu \citep{sasano12,sasano11}.
		(b) Three shell model calculations. 
		Strengths from the 2$^+$ states are also included.
		The vertical lines indicate the region where the KBF model is larger than the GXP model.
		}
	\label{gtni56a_strength}
\end{figure*}

The shell-model calculations are carried out with the code MSHELL \citep{mizu00}, 
allowing at most five nucleons to be excited from the $0f_{7/2}$ orbit. 
The GT strength distributions are obtained by following the prescription
of \cite{white80}.

The GT strength functions for $^{56}$Ni$\rightarrow^{56}$Cu and 
$^{54}$Fe$\rightarrow^{54}$Mn are shown in 
Figure \ref{gtni56a_strength} for several shell model calculations compared
to experimentally determined values \citep{sasano12,sasano11} for $^{56}$Ni.  For the intermediate
excitation energy range at $\sim$4 MeV for the $^{56}$Ni reaction, B(GT) drops in the experiment and
in the GXPF1J model.  However, this decrease does not appear for the KBF
and KB3G models.  The net result is an expected decrease in the EC
rate on $^{56}$Ni.  While the GXPF1J results are not as low at $E_x\sim$4
MeV as the experimental results, the trend in this model is similar to that
of the experimental results at this energy. 

At higher energies ($E_x>$4 MeV), however, the value of B(GT) for the GXPF1J model exceeds those predicted by the KBF and KB3G models. 
This
crossover (indicated by the vertical lines in Figure 
\ref{gtni56a_strength}) above 4 MeV may be less significant as one convolves
the strengths with the Fermi function and the electron energy distribution.
Overall, the EC rates are expected to be lower for the GXPF1J model for a high-enough temperature. 

For the $^{54}$Fe reaction, the drop in B(GT) at $E_x\lesssim$ 4 MeV is less
apparent depending on which KB parametrization is used.  For the KB3G 
model, the the strength function for GXP is lower at $E_x\lesssim$ 3 MeV, and higher above this crossover energy.  The GXP and KBF parametrizations
are similar with a few deviations.

Also shown in Figure \ref{gtni56a_strength} are the strength functions
for electron captures from the first excited state of the parent nucleus. These will be discussed in a later section. 
The EC rates of pf-shell nuclei are evaluated for GXPF1J by including contributions from all the excited states of parent nuclei up to $E_x$ = 2 MeV.  
\begin{figure*}[!]
	\centering
	\includegraphics[width=0.49\textwidth]{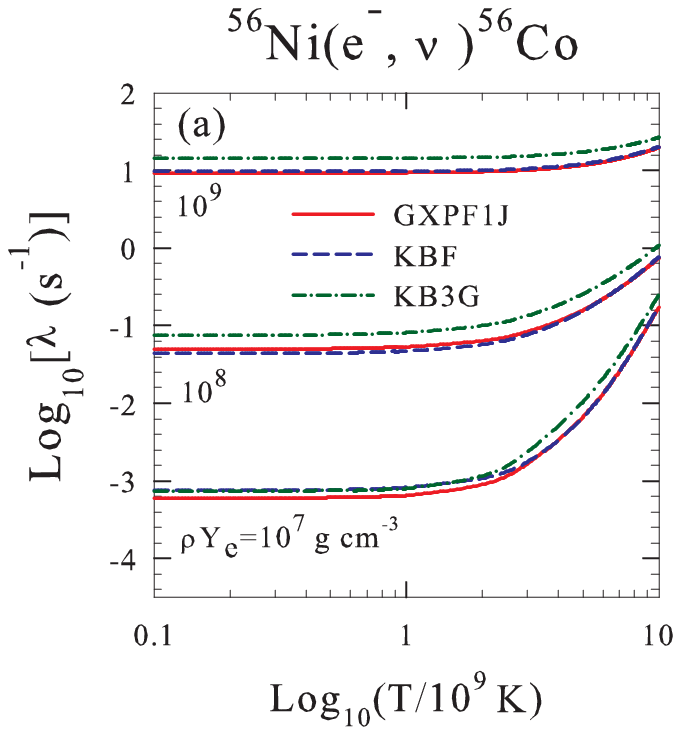}
	\includegraphics[width=0.49\textwidth]{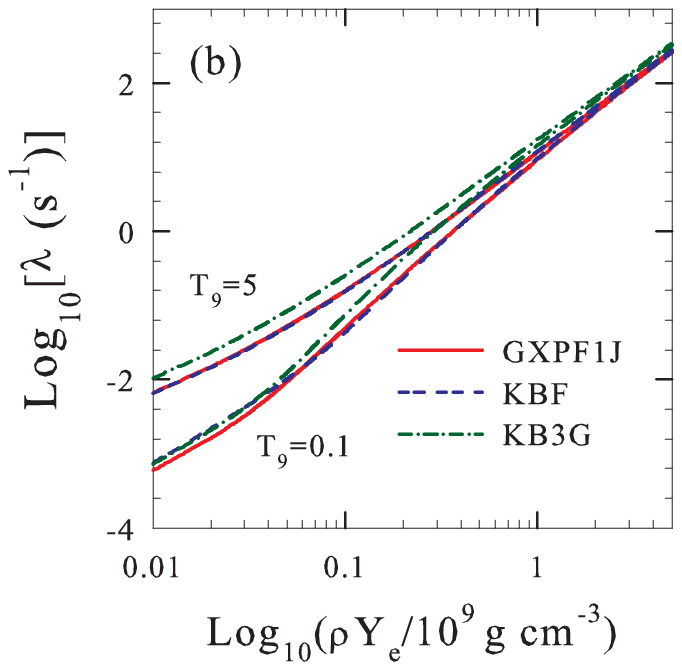}
	\caption{Electron capture rates for $^{56}$Ni as a function of
		(a) temperature for various values of $\rho Y_e$ and (b) $\rho Y_e$
		for various temperatures. Here $T_9\equiv T/(10^9~K$).}
	\label{ni56rates}
\end{figure*}
\begin{figure}[!]
	\includegraphics[width=\linewidth,trim={0.5cm 0 0.5cm 0.3cm}]{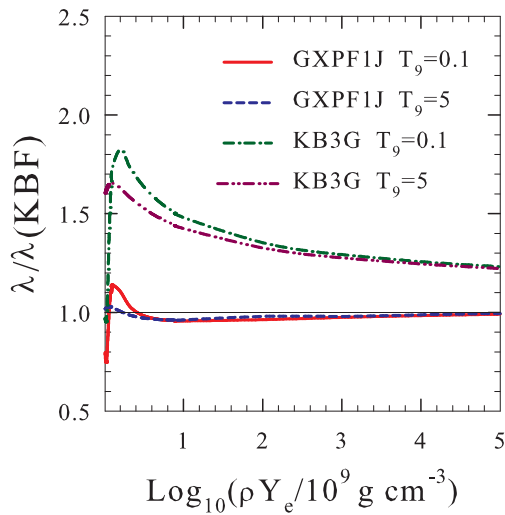}
	\caption{The ratio of the $^{56}$Ni electron capture rate as a function of $\rho Y_e$ for various temperatures.  The ratios for four models
		with respect to the KBF rates \citep{kb3g} are shown. }
	\label{gtni56a_ratio}
\end{figure}

The resultant EC rates for $^{56}$Ni are shown as a function of temperature and $\rho Y_e$ in Figure \ref{ni56rates} comparing the GXPF1J, KBF, and KB3G models. Here $T_9\equiv T/(10^9~K)$.  As expected, the KB3G model produces the
highest rates overall at high temperatures and $\rho Y_e$.  The difference between the KBF and the GXPF1J rates is smaller - a difference
of about 10\% in most cases.  At higher temperatures, the difference between the KBF and GXPF1J rates 
diminishes.  This is likely due to the fact that at higher temperatures, the spread in electron energies
is higher as the Fermi function has a larger effective range.  The integration of the rates over the larger
span in electron energies results in an integration over a larger effective range of energies in B(GT).  
This integration would include the region where B(GT)$_{KBF}>$B(GT)$_{GXPF1J}$ and the region where
B(GT)$_{KBF} <$B(GT)$_{GXPF1J}$.  These two effects would counteract each other.

A comparison of the reaction rates for the $^{56}$Ni EC reaction is shown in Figure \ref{gtni56a_ratio}.  Here, we take the ratio of several rate determinations to those of the KBF rates.  Several shell model calculations for two temperatures as a function of $\rho Y_e$ are employed.  It can be seen that the differences between a chosen shell model and the KBF are largest at low $\rho Y_e$.  It is also noted that at higher temperatures, the rates in all models
converge, as the electron energy distribution is integrated 
over a larger portion of B(GT). However, while the KB3G
rates are larger than the KBF rates at low $Y_e$ by as much as 
$\sim$ 60\%, the enhancement in rates for the GXP model is
only about 10\%, precluding a potentially small effect on
Ni production in SNe Ia.

It is also noted from Figure \ref{gtni56a_strength} that the experimentally determined GT strengths for $^{56}$Ni$\rightarrow ^{56}$Cu exceed those determined using the
	GXPF1J and KBF models slightly for the energy range 1.5 $\lesssim E_x \lesssim$2.5 MeV.  This region may be important for the production of $^{56}$Ni in
	SNIa as the temperature drops significantly at this stage of the nucleosynthesis. A possible increase in EC rates for $^{56}$Ni cannot be ruled out from
	the experimental data.

\begin{figure}
	\includegraphics[width=\linewidth,trim={0.5cm 0.5cm 0.5cm 0.3cm}]{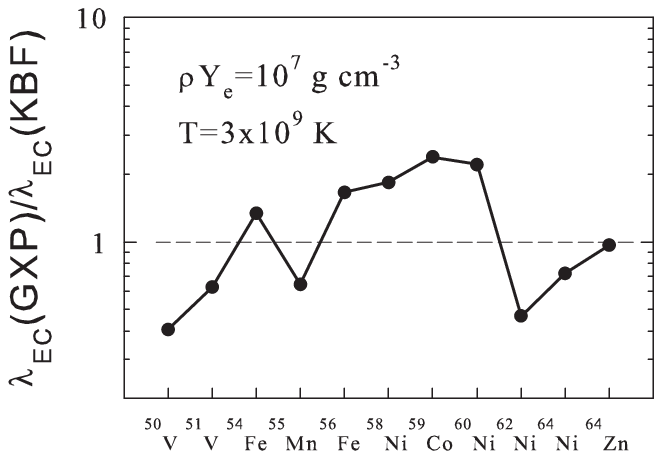}
	\caption{\label{ecapratio}The ratios of electron capture rates from the GXP model to those of the KBF model for 11 nuclei compared in \cite{cole12}.}
\end{figure}

This point is further exemplified in Figure \ref{ecapratio}.  Here, the
	ratio of GXP rates to KBF rates, $\lambda_{GXP}/\lambda_{KBF}$ are shown for
	11 nuclei considered in \cite{cole12}. This evaluation was performed for $T_9=3$ and $\rho Y_e = 10^7$ g~cm$^{-3}$.  From here, no systematic trend is
	determined as the relationship between nuclear structure and the thermodynamic
	environment of a SN Ia is complex. The ratios are within 0.4-2.4, which shows that the e-capture GXP and KBF rates are close to each other - within a factor of 2.5. 
	This is also true for Ga (Z=31) and Ge (Z=32).
	Effects from the use of the GXPF1J model for nuclei with Z=31 and 32 and 23$\le Z \le$30 are, therefore, expected to be modest at best. 
\section{Nucleosynthesis Calculations}
\label{nucleosynthesis}
To evaluate the effects of changes in capture rates on SN Ia nucleosynthesis, the two-dimensional electron-capture (EC) rates 
as a function of $\rho Y_e$ and temperature have been determined based
on the GT strength functions to produce two nuclear reaction networks.  The nuclear reaction networks used are 
	summarized in Table \ref{network_list}.  Two different rate tables were used in combination with two different SN Ia explosion
	scenarios modeled by their hydrodynamic trajectories.
\begin{table*}
	\caption{\label{network_list}Summary of the nuclear reaction networks used in this study along with the rate tables 
		and hydrodynamic explosion trajectories used \citep{nomoto84,nomoto}.}
	\begin{center}
		\begin{tabular}{|l|c|c|c|}
			\hline
			\textbf{Model} & \textbf{Explosion} &\textbf{$\lambda(Z<21)$} & \textbf{$\lambda(21\le Z \le 32)$}
			\\\hline
			I & W7 & \cite{ffn1,ffn2,oda} & KBF \\\hline
			II & W7 & \cite{ffn1,ffn2,oda} & GXP \\\hline
			III & WDD2 & \cite{ffn1,ffn2,oda} & KBF \\\hline
			IV & WDD2 & \cite{ffn1,ffn2,oda} & GXP \\\hline
		\end{tabular}
	\end{center}
\end{table*}

The first network employed EC rates for the KBF model \citep{kbf,langanke01} for proton-rich pf-shell nuclei with 21$\le$Z$\le$32 (for mass A, 45$\le A \le$ 65).  The second network
used rates computed from the GXP shell model for
the same subset of nuclei. In both networks, for nuclei outside this region, the rates of \cite{oda}
were used for $sd$-shell nuclei while the rates of
\cite{ffn1,ffn2} were used otherwise. All other reaction rates were taken from the JINA REACLIB database \citep{reaclib}.  The {\it libnucnet} 
	reaction network engine was used for the nucleosynthesis calculations \citep{libnucnet}.

Both nuclear reaction networks were run using the hydrodynamics of the W7 deflagration \citep{nomoto84} and WDD2 delayed-detonation 
WD explosion model \citep{nomoto}. Trajectories for the deflagration and shock-front burning were
followed.   From this, an analysis and comparison of the production of pf-shell nuclei were done similar to that performed 
previously \citep{brachwitz00,nomoto}.

In this evaluation, the individual trajectories of the explosion models are 
decoupled from the nucleosynthesis and used as inputs in the nuclear reaction
network.  Each trajectory is a mass layer in the explosion.  The electron chemical 
potential and electron fraction are computed implicitly at each time step for each 
trajectory and used as inputs in the weak rates. It is noted that while decoupling the
	reaction network from the explosion trajectories allows for a rapid evaluation of the effects of multiple shell models on the nucleosynthesis, the differences in heating induced by differences in 
	the reaction rates is not accounted for.  The uncertainty in this approximation will be discussed in the
	next section.
\section{Results}
\label{results}
Nuclear reaction network calculations were run for the central trajectories in the W7 deflagration and the WDD2 delayed-detonation 
explosion models \citep{nomoto}. The final abundances
were computed based on these network calculations.  Rates for pf-shell nuclei were computed for both the GXPF1J shell model and
the KBF model \citep{langanke01}.  (Here, the reaction network results are referred to as the GXP and KBF models for networks using the GXPF1J and KBF 
shell models, respectively.)  
\subsection{Deflagration Model}
As mentioned in \S\ref{intro} deflagration models, such as the W7 model\citep{nomoto84,thielemann86}, attempt to simulate
	the effects of increased nuclear burning because of
	an increase in the surface area of the flame front as
	it propagates to the stellar surface.
	The flame (deflagration) speed is prescribed by mixing-length theory.
	It accelerates to 0.08$c_s$ in t=0.6 sec and to 0.3$c_s$ in t=1.18 sec, where $c_s$ is the local sound speed
	\citep{nomoto84}.

\begin{figure}[h]
	\includegraphics[width=\linewidth]{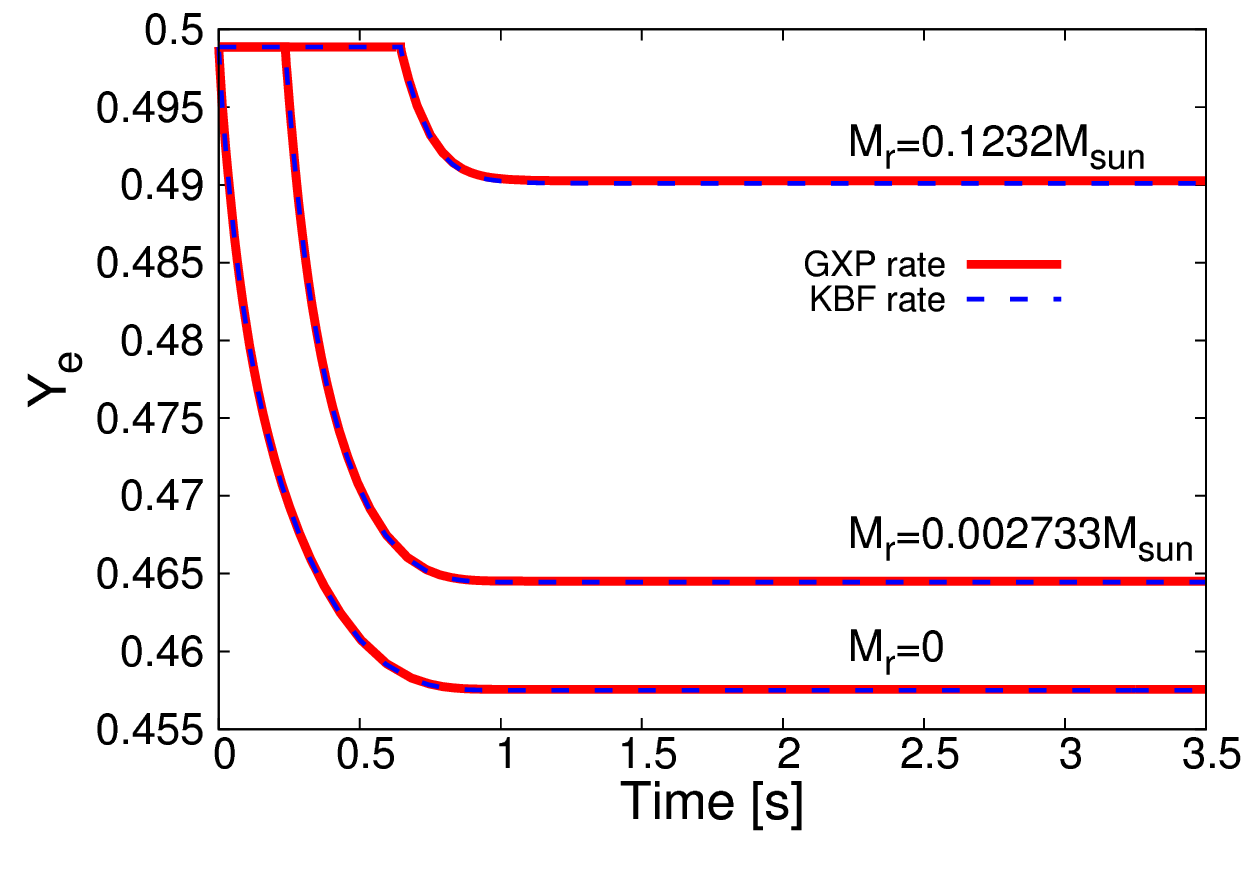}
	\caption{\label{ye_time}The evolution of the electron fraction for several trajectories in the GXP model (model II), given by
		the solid lines, and the KBF model (model I), given by the dashed line.}
\end{figure}

The deflagration model W7 was explored using networks I and II in Table \ref{network_list}. Here, the
	thermodynamic trajectories of \cite{nomoto84} have
	been used as inputs to the nuclear reaction network.
The evolution of the electron fraction $Y_e$ is shown in Figure \ref{ye_time} for several trajectories in both models.
The values of $Y_e$ at the end of the nucleosynthesis calculations ($t\approx 3.5$ s) are also shown in Table 
\ref{fin_ye} for the same trajectories.  In this figure, each
line represents the evolution of a mass element for the W7 model assuming
a specific shell model assumption. 
For all models, the electron fraction is lower near the core.  It can
be seen that the electron fraction changes little between the GXP and the KBF models indicating that, though the GXP shell model results in a 
significantly different B(GT), the net effects on the nucleosynthesis are
potentially insignificant.
\begin{table}[h]
	\caption{\label{fin_ye}The electron fractions at 3.5 s for the trajectories in Figure \ref{ye_time} for each shell model used in this paper.
		Roman numerals in parentheses indicate the model number indicated in Table \ref{network_list}.}
	\begin{tabular}{|c||c|c|c|}
		\hline
		~&$M_r$=0&$M_r/M_\odot$=&$M_r/M_\odot$=\\
		~&~& 0.002733& 0.1232\\\hline\hline
		GXP (II)&0.457555&0.464515&0.490261\\
		\hline
		KBF (I)&0.457501&0.464453&0.490101\\
		\hline
	\end{tabular}
\end{table}  

\begin{figure*}[!]
	\centering
	\includegraphics[width=0.49\textwidth]{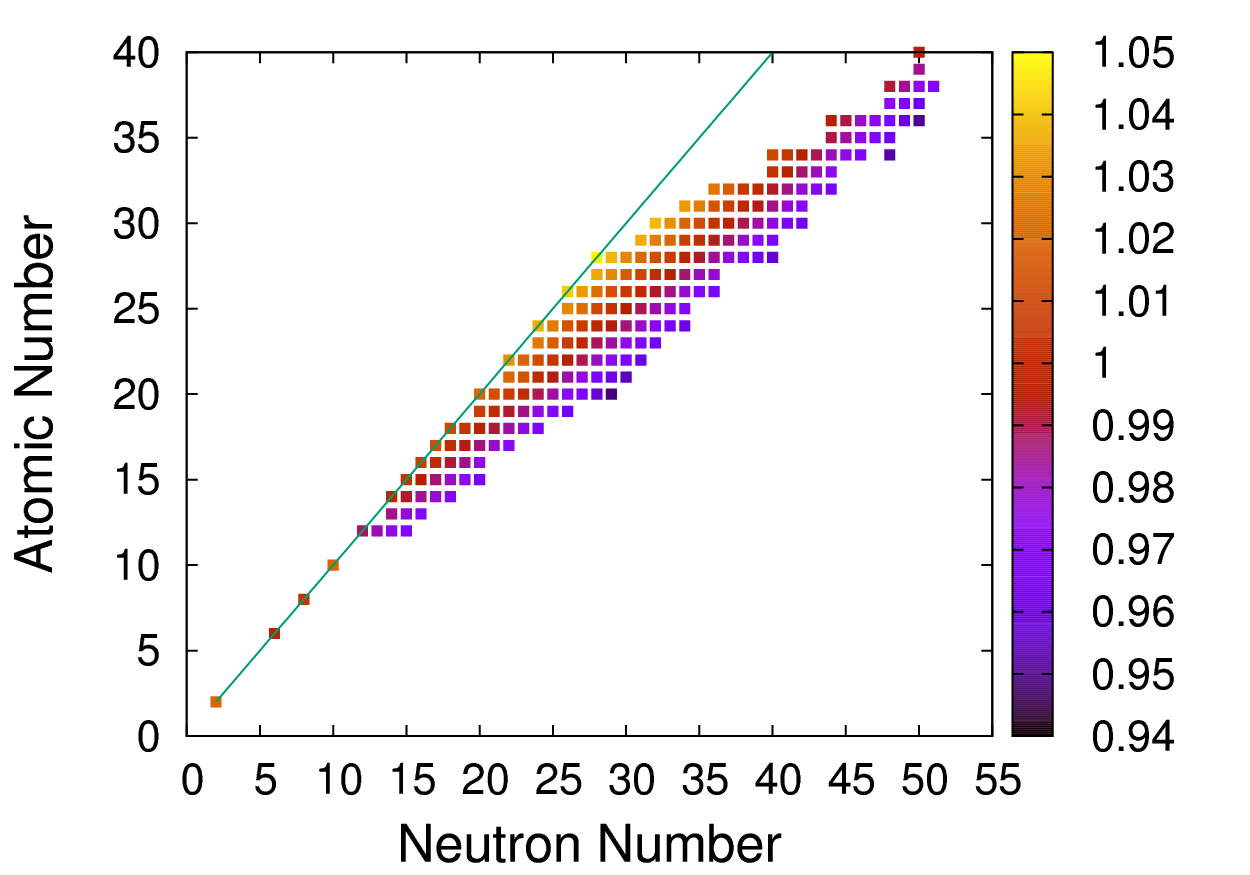}
	\includegraphics[width=0.49\textwidth]{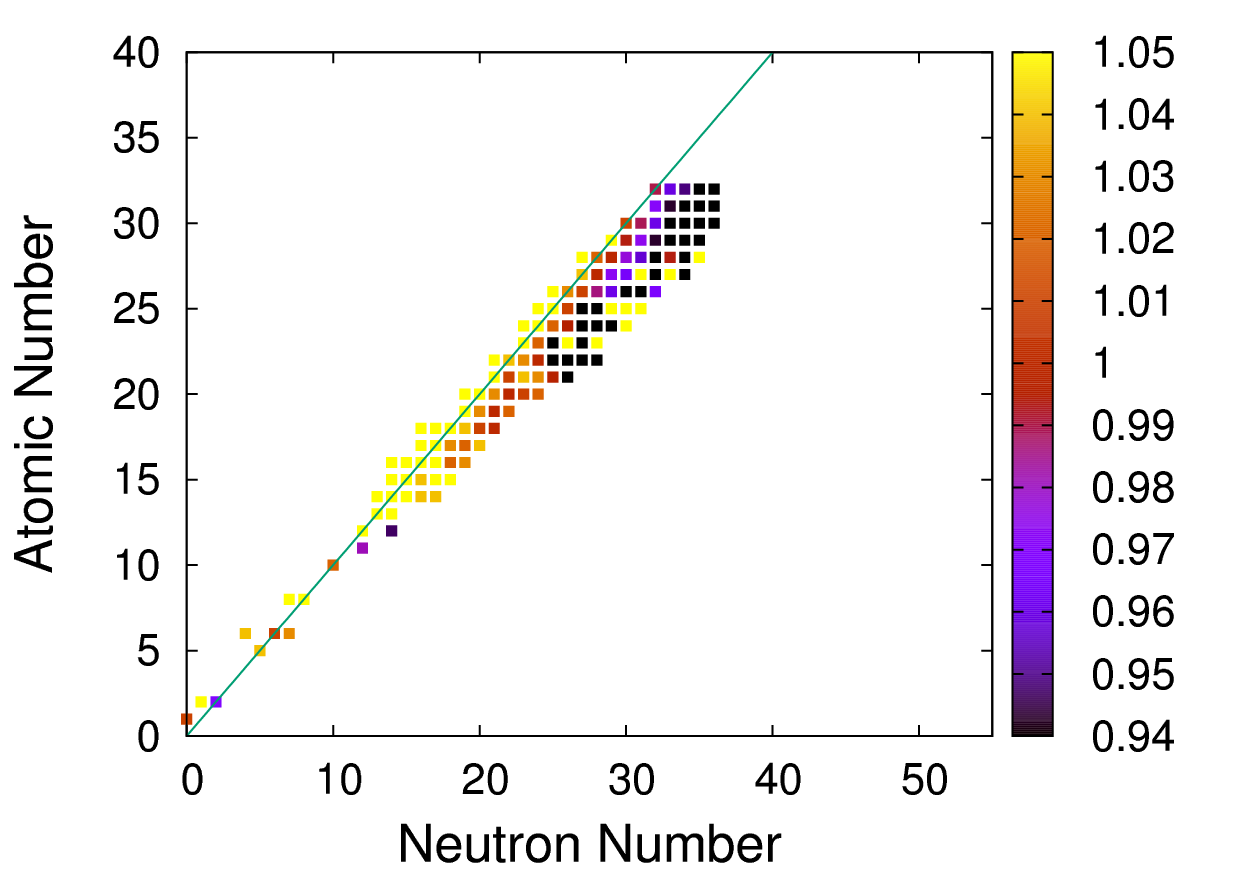}
	\caption{Abundance ratios $Y_{GXP}/Y_{KBF}$ at $t=3.5$s in the core, $M_r$=0 (left) and for the mass element located at 
		$M_r$=0.1232$M_\odot$ (right) showing an abundance shift towards 
		more proton-rich nuclei. The N=Z line is indicated in the plots.}
	\label{ratio1}
\end{figure*}

Figure \ref{ratio1} shows the final mass fraction ratios for both calculations:
\begin{equation}
R\equiv \frac{X(Z,N)_{GXP}}{X(Z,N)_{KBF}}
\end{equation}
Ratios are shown for the central trajectory (at a mass radius $M_r$=0) and an off-center trajectory (at $M_r$=0.1232$M_\odot$).  It can be seen that the GXP model
results in a more proton-rich final abundance distribution, but the effect is small. While this is not surprising as the electron capture rates are lower in the pf-shell region for the GXP model, the overall increase in  $Y_e$ (as seen in Figure \ref{ye_time})
is small.

It is also noted that the final abundances for the GXP model are more proton-rich for nuclei outside of the region where the rates
differ.  Of course, a reduction in the EC rates results in a global increase in $Y_e$, which will affect all electron capture
rates.  Thus, while the overall electron fraction may change only slightly,
the final abundance distriubution for a particular species may
change by a larger amount.

\begin{figure}[!]
	\includegraphics[width=\linewidth]{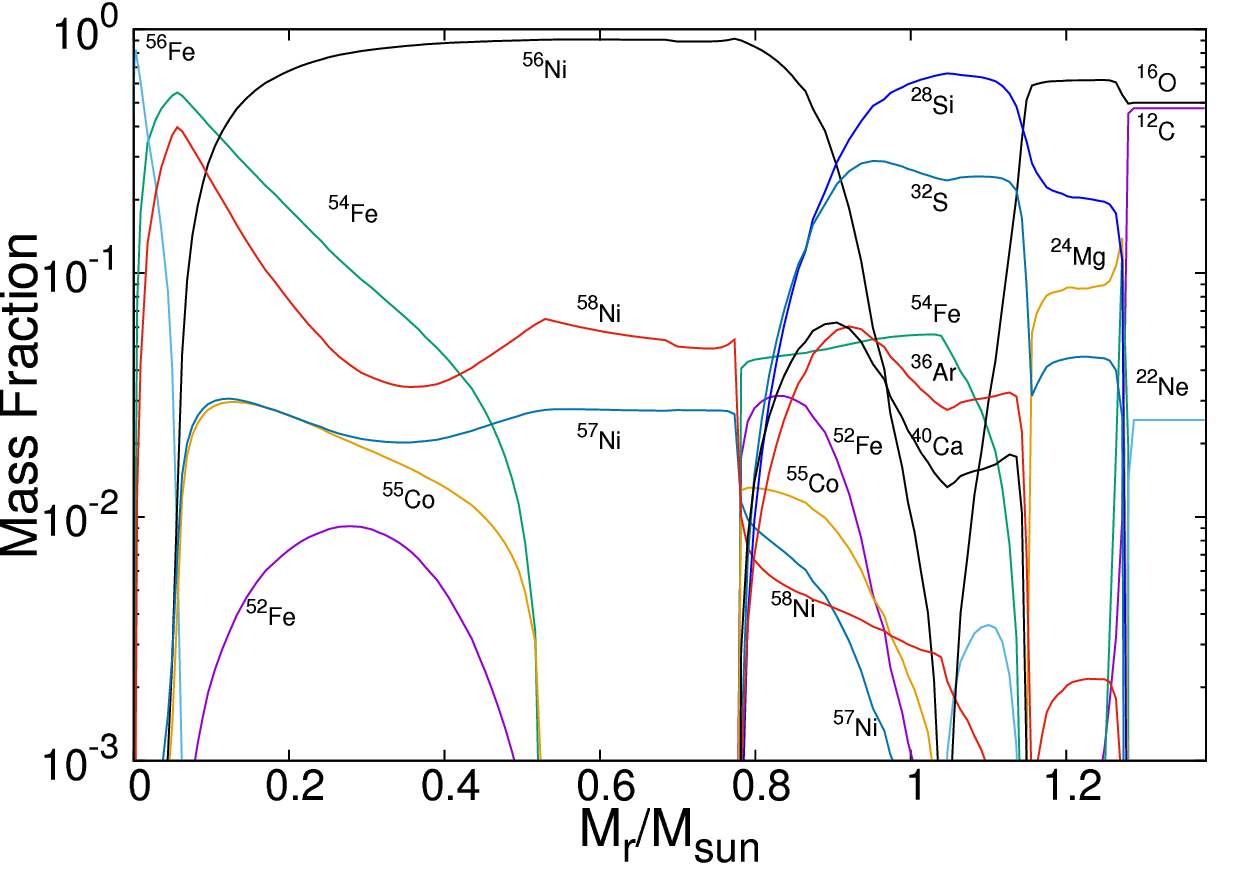}
	\caption{\label{final_masses}Isotopic distributions
		as a function of mass using the GXP rates (model II) at the end of the nucleosynthesis calculations($t=3.5$ s).  
		The results for the KBF model are nearly identical.} 
\end{figure}

Using the final abundances for each trajectory in the W7 model, the final 
mass fraction profile has been determined.  These profiles are shown in
Figure \ref{final_masses} for the GXP  model evaluated in this work.  These mass fractions closely resemble those of prior 
work \citep{brachwitz00}.

\begin{figure}[!]
	\centering
	\includegraphics[width=\linewidth]{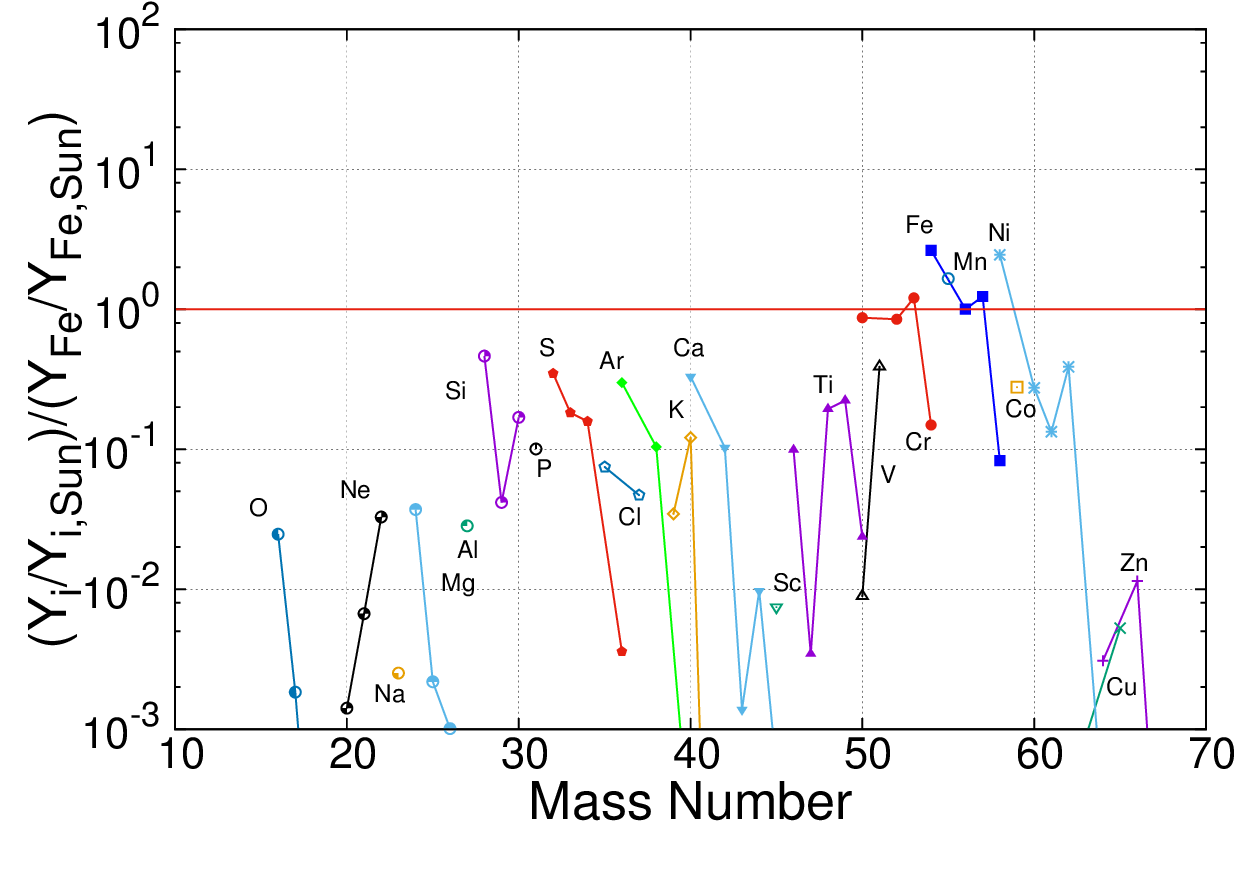}
	\caption{\label{weighted_ratio}Deflagration (W7) final abundance ratios relative to Fe relative to the solar 
		abundance ratio. The ratio is for the GXP model (model II), and results for the KBF model (model I) are nearly identical.}
\end{figure}

From the final abundance ratios in each trajectory, a total abundance 
weighted by the mass of each trajectory
is computed for the isotopes in the network - the overproduction factor.  This double ratio, explicitly defined as:
\begin{equation}
DR\equiv\frac{{Y_i}/{Y_{Fe}}}{{Y_{i,\odot}}/{Y_{Fe,\odot}}}
=\frac{Y_i/Y_{i,\odot}}{Y_{Fe}/Y_{Fe,\odot}}
\end{equation}
provides a measure of the uniformity of the nucleosynthesis compared to solar observations.  
These double ratios are shown in Figure \ref{weighted_ratio}
for the GXP model.  The ratios for the KBF model are nearly
identical and comparable to the results of \cite{brachwitz00}.  Notable are the overabundances of $^{58}$Ni compared to the production of lighter Z nuclei.

The nuclei $^{54}$Mn and $^{54}$Cr can also be addressed in Figure \ref{weighted_ratio}.  Although a small fraction of
$^{54}$Cr is created from electron captures on $^{54}$Mn, $^{54}$Cr is produced in 
significant abundance in the inner 10$^{-3}M_\odot$ of the model, where the electron fraction $Y_e<$ 0.46. Thus, while 
it is not seen in Figure \ref{final_masses} because the scaling of this figure is not fine enough,
its production is still evident in Figure \ref{weighted_ratio}.  It is not surprising that $^{54}$Cr is produced in greatest
abundance in the center mass shell as the $Y_e$ of this shell reaches a value very close to that of $^{54}$Cr ($Y_e\sim Z/A=0.44$) as seen from 
Figure \ref{ye_time} and Table \ref{fin_ye}.  From a 
nuclear statistical equilibrium (NSE) argument, one would expect this to be the mass shell with the largest final
abundance of $^{54}$Cr.  This is also the mass shell with the lowest final $Y_e$.  It is possible that
an overall increase in $Y_e$ of the entire model by a small amount could result in dramatic changes in the $^{54}$Cr production. This is because all of the mass shells above the central shell would have $Y_e$ which is even farther away from that of $^{54}$Cr.  If the $Y_e$ of the central shell increases to above 0.46 (a shift of less than 1\%), then the overall $^{54}$Cr production would start to 
decrease.  However, production of nuclei with a higher $Y_e$, such as $^{56}$Ni, would not be expected to decrease significantly because it is produced in a large range of 
mass shells with a range of electron fractions above and below that of $^{56}$Ni ($Y_e\approx Z/A = 0.5$).  In the case of $^{56}$Ni, while a single shell closer to the surface may produce less $^{56}$Ni, shells below it would have an
increased production, thus compensating for the loss.  The inner mass shell has no shells below it, and any loss suffered 
by an overall increase in $Y_e$  could not be compensated for.  Even more, a global shift of $Y_e$ for all shells may also increase the overall production of $\alpha$-cluster elements such as $^{28}$Si.  

For this reason, $^{54}$Cr could be a good indicator of the effectiveness of the nucleosynthesis in a SN Ia as it is produced 
predominately in a small region of the star.  Small global shifts, which tend to be averaged into multiple regions of the 
star, have more significant effects for the central region.  Because of this, it is worthwhile to explore nucleosynthesis and thermodynamic effects such as turbulence and mixing within the SN Ia.  In this case, 3D modeling - or 1D and 2D approximations -  would become important.

\begin{figure}[!]
	\includegraphics[width=\linewidth]{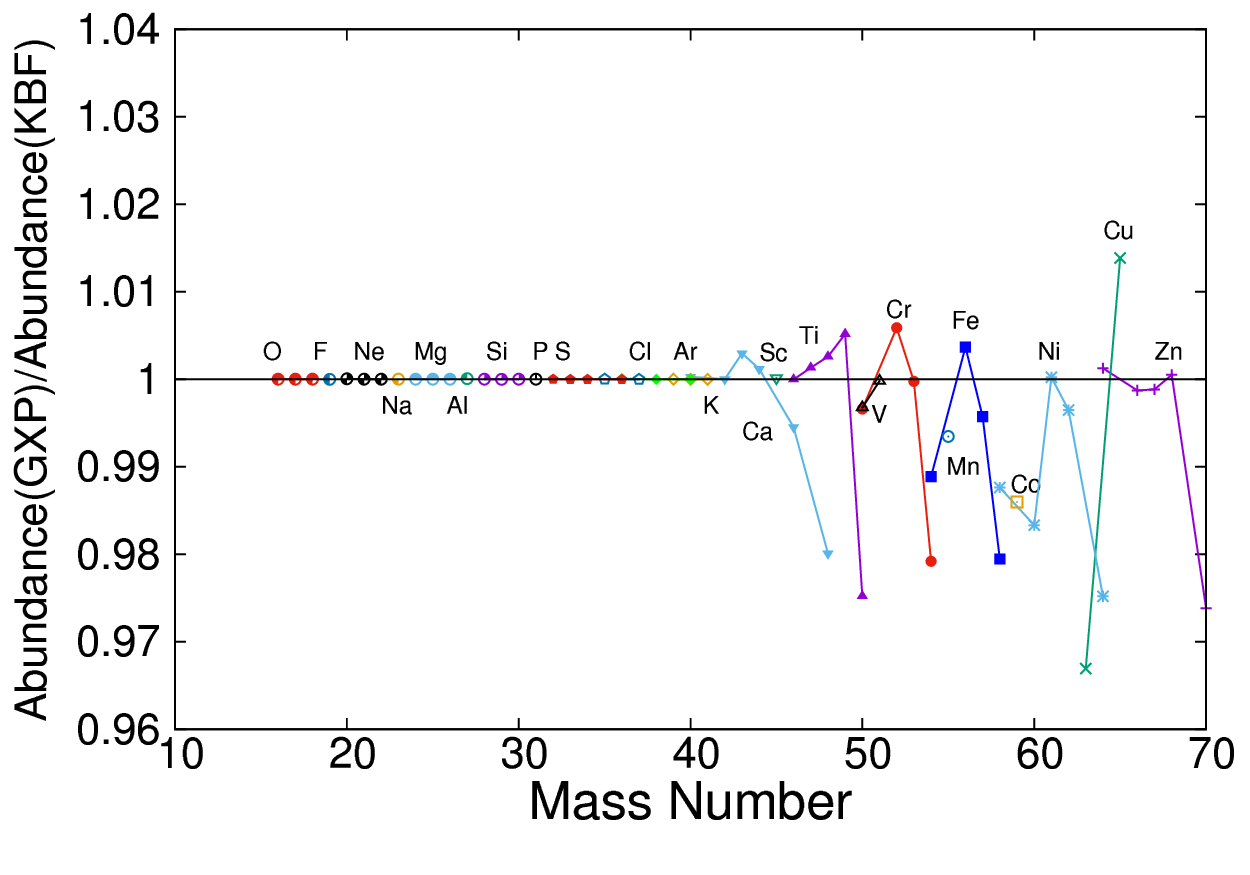}
	\caption{\label{ratio_abun}The ratio of abundances for nuclei produced in the GXP model (model II) to those of nuclei produced in the KBF model (model I).  
		The results shown here are for the W7 deflagration model \citep{nomoto84}.}
\end{figure}

To compare the GXP and KBF models, we take a ratio of the abundances produced in each model weighted by the individual mass shells in the 
explosion. This ratio is shown in Figure \ref{ratio_abun}. A small reduction
of $\lesssim$5\% is noted for the pf-shell nuclei with a slight increase in only a few cases.  The production generally decreases
in the GXP model for higher neutron number, consistent with a slight
shift to higher $Y_e$ in the GXP model.  

Despite the seemingly significant differences in B(GT) between the GXP and KBF
models, why are the differences in the nucleosynthesis so small (and likely unobservable)?  The answer lies in the intersection of nuclear physics and the hydrodynamics of the deflagration model.

The total electron capture rate for a transition from a parent state to 
a daughter state is:
\begin{align}
\label{ec_rate}
\lambda=&\ln2 \frac{(g_A/g_V)^2}{6143}\sum_i \frac{\left(2J_i+1\right)e^{-E_i/kT}}{G(Z,A,T)}\times
\\\nonumber
&\sum_j{B_{ij}(GT) f_{ij}(T,\rho,\mu)}
\end{align}
where the sum is over transitions from parent to daughter states.  The
phase space factor, $f_{ij}$, accounts for the energetic feasibility of the reaction,
including the population of electrons at a particular energy \citep{martinez00}:
\begin{align}
\label{phase_space}
f_{ij} &= \int_{\omega_l}^{\infty}\omega p\left(Q_{ij}+\omega 
\right)^2 F(Z,\omega)S_e(T)\\
\nonumber
~&\times (1-S_\nu(Q_{ij}+\omega)) d\omega
\end{align}
where $Q_{ij}$ is the electron capture transition energy divided by the electron mass, determined from 
the nuclear masses, and transitions are summed in Equation \ref{ec_rate} from
initial states $E_i$ to final states $E_j$.  The distribution function is $S_{e,\nu}(\omega)$; 
here, $S_e$ is the usual Fermi-Dirac distribution, and $S_\nu=0$. The normalized total electron energy (kinetic plus 
rest mass) is $\omega = E_e/m_ec^2$.  Coulomb
barrier penetration effects are contained within the function $F(Z,\omega)$. The integration limits are set by the reaction threshold
$\omega_l$, which is the threshold value for which EC is energetically
feasible; $\omega_l=1$ if $Q_{ij}>-1$ and $\omega_l=\left|Q_{ij}\right|$ for $Q_{ij}<-1$. 
	For $Q_{ij}<-1$, the electron must have enough
kinetic energy to make the reaction possible, and $\omega_l>1$.  An electron with zero kinetic energy
will not exceed the reaction threshold.  

We consider the case of electron captures on $^{54}$Fe.
The ground-state to ground-state Q-value for $^{54}$Fe EC (using nuclear
masses) is $-$1.21 MeV, meaning that to integrate over the 
strength function shown in Figure \ref{gtni56a_strength},
the electron must have a total energy of at least 1.21 MeV. This raises
the integration threshold in Equation \ref{phase_space}.   In order
for the Fermi distribution $S_e(T)$ to have a population with the electron
energy above the threshold $\omega>\omega_l$, either the density must
be high enough to push the Fermi energy above the threshold, or the temperature
must be hot enough so that the smearing of the Fermi surface creates
a population of electrons above threshold, or both.  

\begin{figure*}[!]
	\centering
	\includegraphics[width=0.49\textwidth]{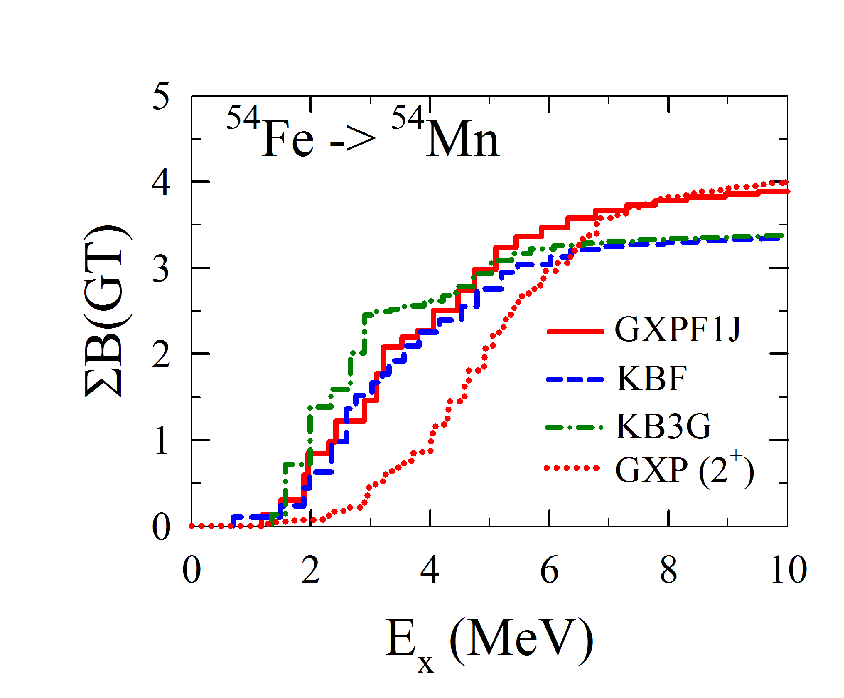}
	\includegraphics[width=0.49\textwidth]{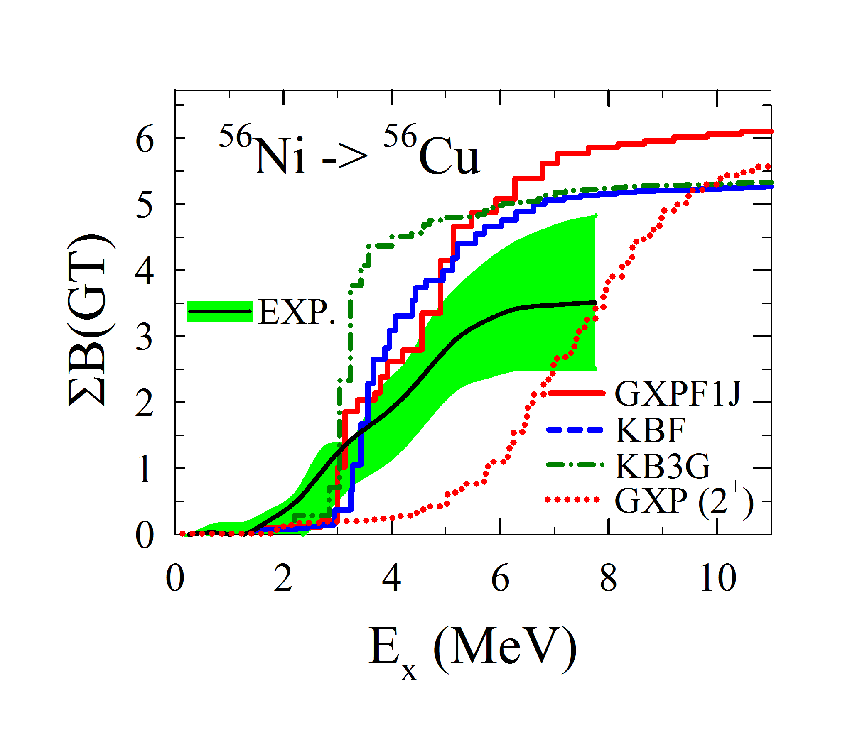}

	\caption{Integrated GT strengths vs. excitation energy for the three models compared in this paper. {\bf Left: }$^{54}$Fe.
		{\bf Right: }$^{56}$Ni along with experimental values.  Also shown in each figure are the 
		sums for EC transitions from the first excited
		state in the parent nuclei.}
	\label{sumgt}
\end{figure*}

However, during the nucleosynthesis, the time at which the production of
$^{54}$Fe is reached does not occur until about 0.5 s in the mass shell which has the largest mass fraction
of Fe by the end of the nucleosynthesis, corresponding to a mass of
$5.612\times 10^{-2} M_\odot$.  By this time, the temperature has dropped significantly in
the W7 model, and the Fermi energy is less than 1.5 MeV. The spread
in the Fermi surface is roughly 1 MeV, so nearly all electrons have energies
less than 2 MeV.  This means that the sum over B(GT) in Figure \ref{gtni56a_strength} is only at excitation energies less than 2 MeV. One
sees from Figure \ref{gtni56a_strength} that the differences between the KBF
and GXP strength functions are small in this region.  This is indicated in
Figure \ref{sumgt} which shows the integrated strength functions
vs. the excitation energy of the daughter nucleus.  For electron
captures on $^{54}$Fe, only excitation energies less than 2 MeV are 
important, in a region where the KBF and GXP strengths are nearly equal.
\begin{figure*}
	\includegraphics[width=0.49\linewidth]{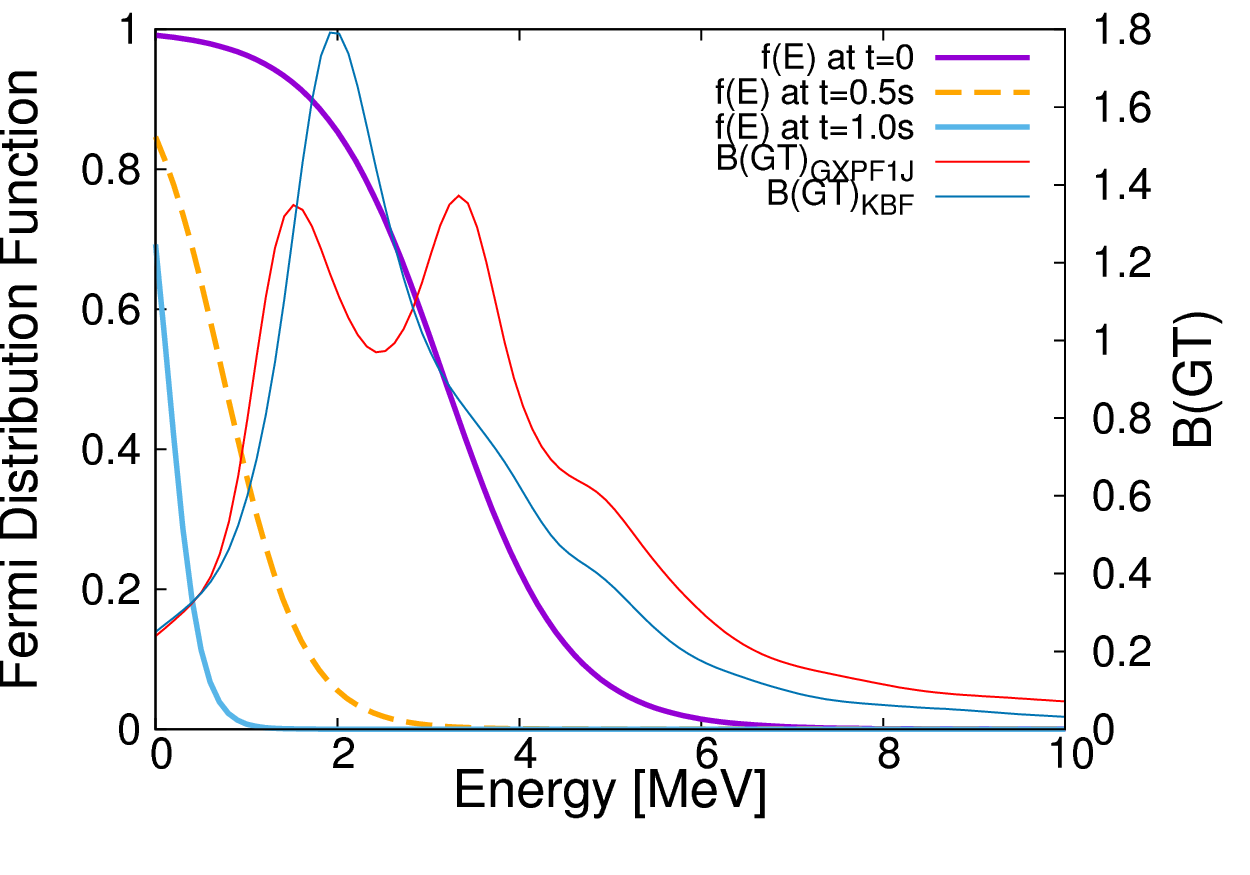}
	\includegraphics[width=0.49\linewidth]{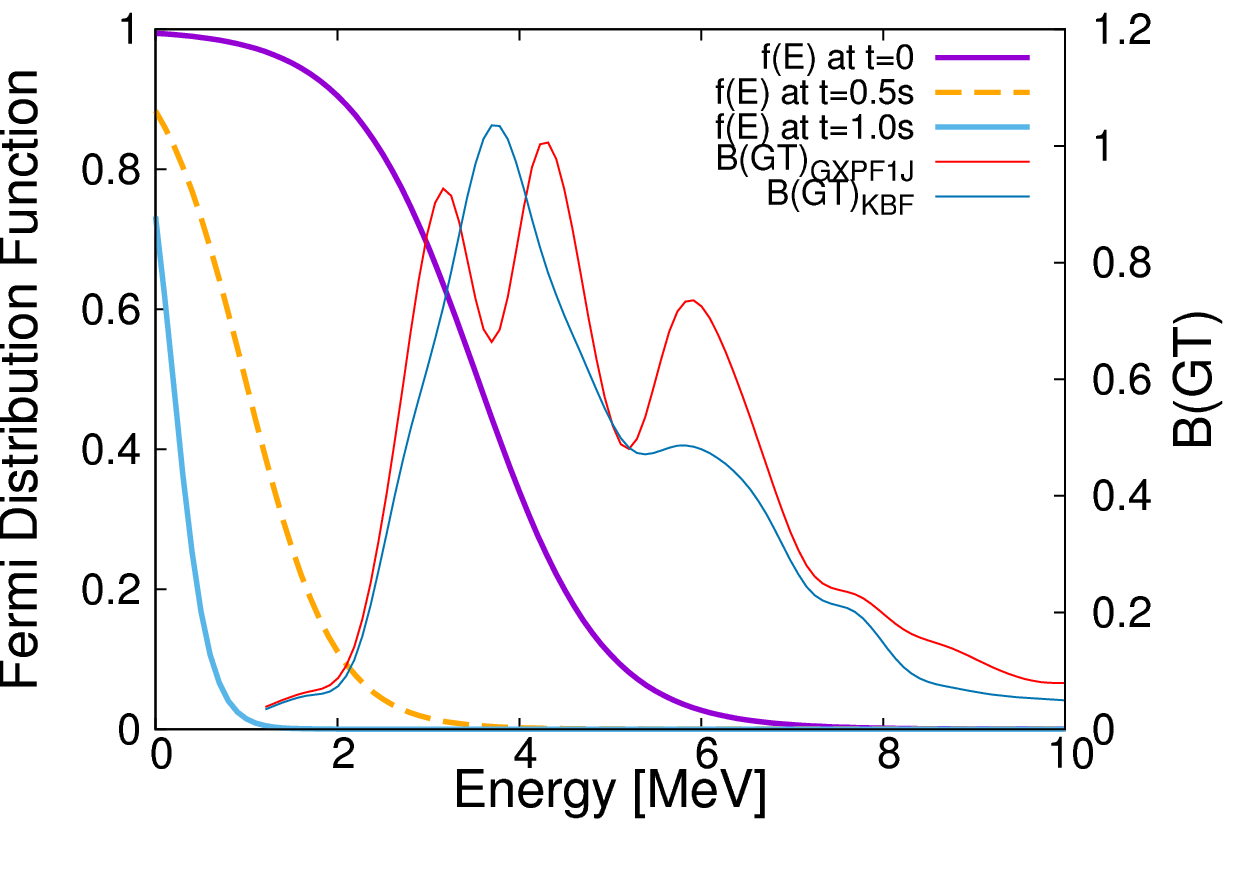}
	\caption{Fermi distributions for mass shells that produce the largest amount of $^{56}$Ni (left) and $^{54}$Fe (right) for three different times in the mass shell evolution.  The distributions are shown with the B(GT) strengths for both nuclei.  The B(GT) strengths have been shifted by the EC Q-value to show the electron energy necessary from each Fermi distribution to populate the specific excitation energy in the strength function.
		The thermodynamic trajectories are those of the deflagration model W7, corresponding to models I and II in this work.\label{evolution}}
\end{figure*}

The case is similar for electron captures on $^{56}$Ni. Although the Q-value
is positive for this reaction (1.622 MeV), the temperature and density by the time 
$^{56}$Ni is reached in the nucleosynthesis are roughly the same as for
$^{54}$Fe.  While a small population of electron energies exceeds 3 MeV
where there is some deviation between the GXP and KBF models, most of the electron population exists at energies less than 1 MeV, corresponding to
excitation energies less than 2.6 MeV, where the electron capture rates are
low and the difference between the GXP and KBF strength functions is negligible.
This can be seen by examining the integrated strength functions in Figure \ref{sumgt} in which deviations between the GXP and KBF models do not occur until excitation energies greater than about 3 MeV for $^{56}$Ni with less
deviation for $^{54}$Fe except at very high excitation energies.

This is shown schematically in Figure \ref{evolution}. In this
figure the GT strengths are shown for $^{56}$Ni and $^{54}$Fe for the KBF and GXPF1J shell models.  Electron Fermi distributions for the mass shells that produce the largest amount of $^{56}$Ni and $^{54}$Fe are overlayed onto these figures for various times along the mass shell evolution.  The
GT strengths have been offset by the electron capture Q-value (ground state) in each case to provide an accurate picture of the actual electron energy necessary to reach a specific excited state in the
daughter nucleus.  

It is seen from this figure that while the mass shell starts off hot enough to create a population of electrons energetic
enough to populate parts of the GT strength where
the differences are significant between the two shell models,
one must keep in mind that the nucleosynthesis path for each mass shell does not reach $^{56}$Ni and $^{54}$Fe until later
in the shell evolution, in both cases about 0.5 s after the explosion starts.  By this time, the mass shell has cooled to the point at which nearly the entire Fermi distribution lies in the energy region where the difference between the GXP and KBF strengths is insignificant as shown in the figure.  Thus, the difference in EC rates for these two nuclei is small for these explosion models.

From Figures \ref{weighted_ratio} and \ref{ratio_abun}, it can be seen that the difference between the abundances determined using the GXP model and the KBF model are very small - on the order of a few percent.
Furthermore, it can be seen that the abundance differences only exist at Z$>$19.  By the time these nuclei
are produced, the temperature of the environment is low and the nuclear abundances match those very close to an environment in NSE.  Any heating from nucleosynthesis for Z$>$19 is not expected to differ significantly
from that of the model of \cite{nomoto}.  Thus, the decoupling of the reaction network from the explosion trajectories is not expected to produce significant uncertainties.
\subsection{Delayed-Detonation Model}
Nucleosynthesis calculations were also carried out for the 
delayed-detonation model \citep{nomoto} using models III and IV in Table \ref{network_list}.  
In this model,
the explosion transitions from a 
deflagration near its center to a detonation at low density
\citep{khokhlov91}.
This particular model is parametrized by the transition
density with the transition density parameter set to match observed light curves and nucleosynthesis. 

In the delayed detonation, a slow deflagration phase is calculated by \cite{nomoto84}
with an assumed constant flame speed of 0.015$c_s$.
This is a typical laminar deflagration speed without convection and flame instabilities.
The subsequent detonation (shock) phase is calculated by 
\cite{nomoto}.
It is assumed that DDT happens  when the density at the flame front (upstream density) decreases to 2.2$\times$10$^7$ 
g~cm$^{-3}$.

Detailed nucleosynthesis calculations of delayed detonation models find that the problem of overproduction of neutron-rich isotopes may be remedied    
by constraining the density at which DDT occurs to \(\sim 10^7\) g cm\(^{-3}\)
\citep{nomoto,brachwitz00},
which agrees with more physically based estimates 
\citep{bychkov95,niemeyer97,woosley07}.  
This is consistent with the results of multi-dimensional simulations of delayed detonation models 
\citep{seitenzahl13}.

\begin{figure}[!]
	\includegraphics[width=\linewidth]{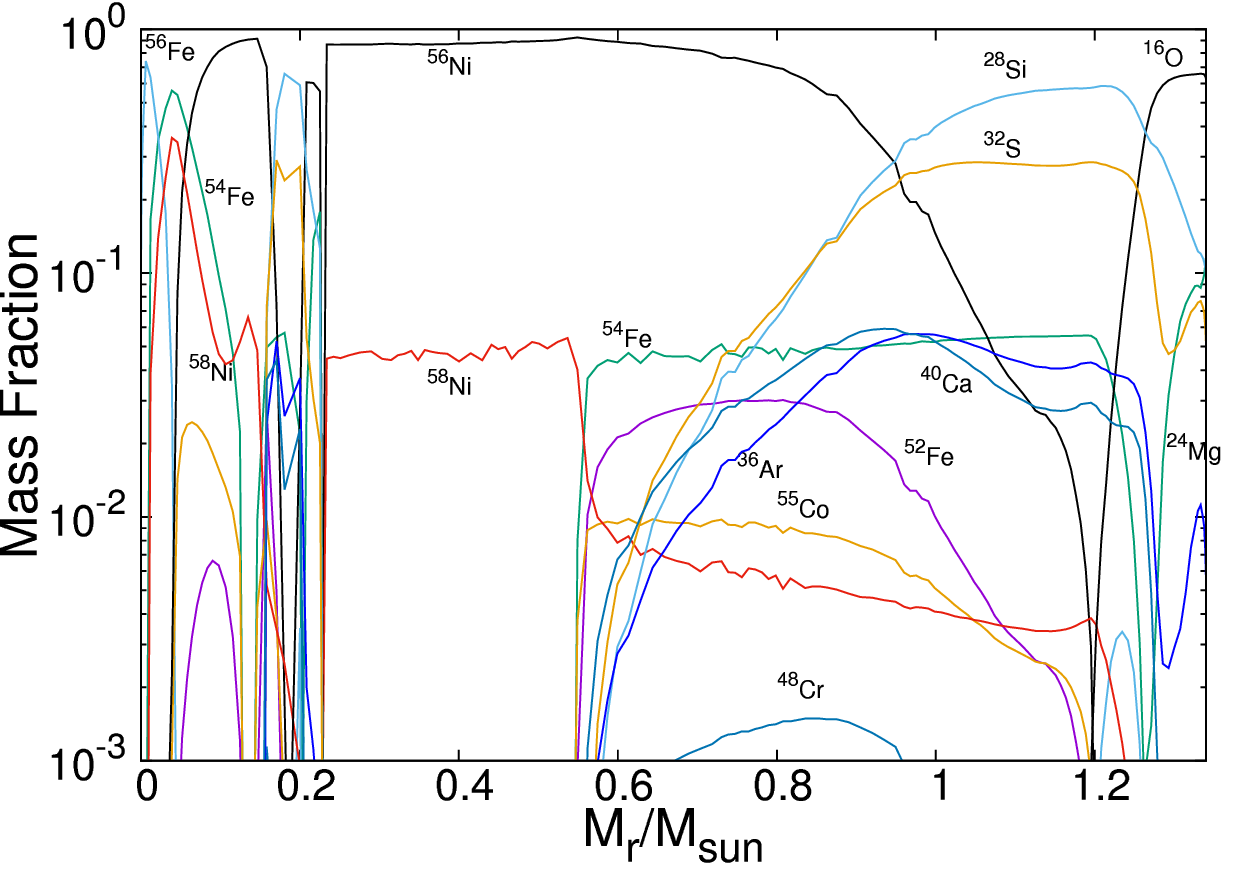}
	\caption{\label{wdd2_massfrac}Same as Figure \ref{final_masses}, but for the WDD2 delayed-detonation model.}
\end{figure}

As with the deflagration models (I and II), calculations were done assuming both a GXP and KBF shell model.   
The resulting mass fractions using the WDD2 hydrodynamic trajectories are shown in Figure \ref{wdd2_massfrac}.  These mass fractions compare 
	to those found using the W7 model of \cite{nomoto}.

\begin{figure}[!]
	\includegraphics[width=\linewidth]{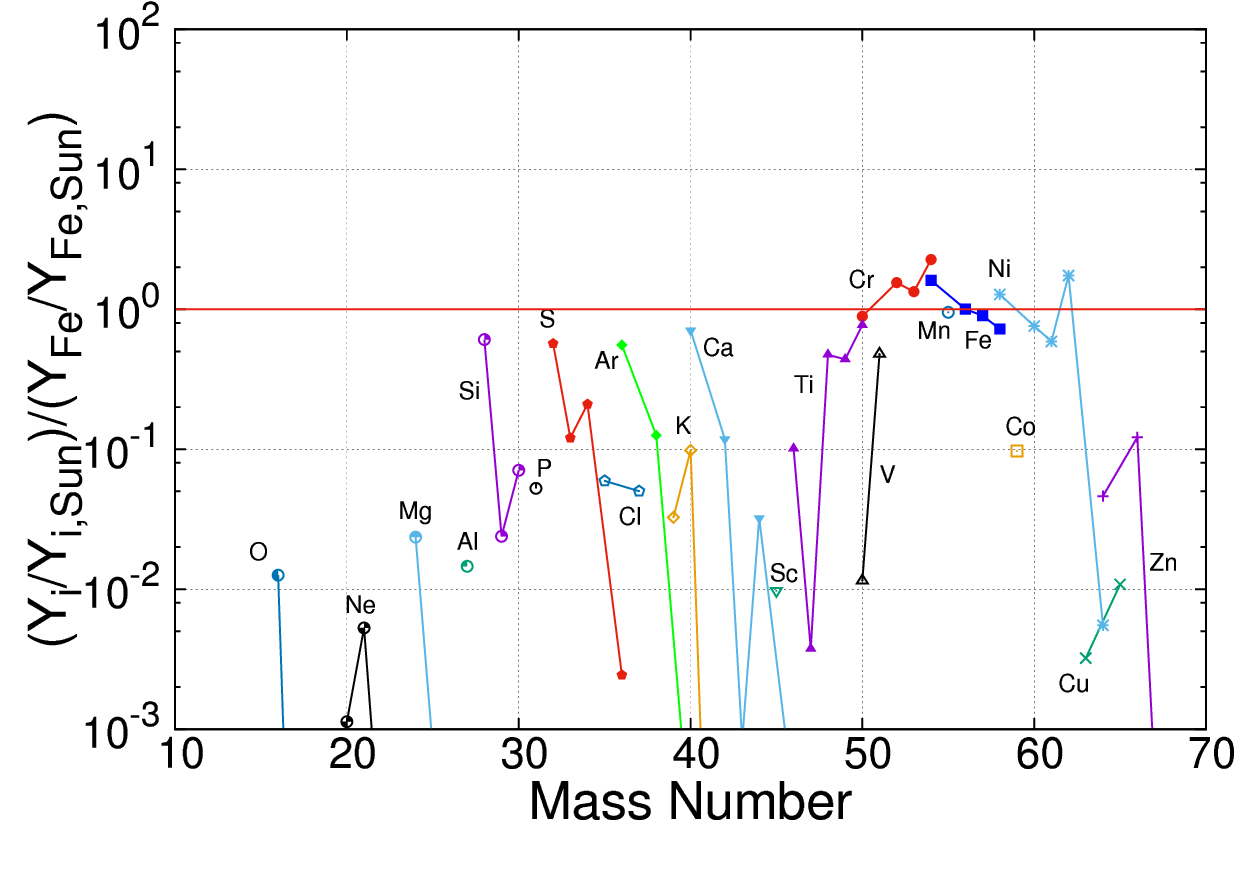}
	\caption{\label{wdd2_gxp}The ratio of abundances for nuclei produced in the GXP model for
		the WDD2 delayed-detonation model corresponding to model IV in Table \ref{network_list}.}
\end{figure}

Nuclear overproduction factors for nucleosynthesis using the  GXP model (model IV) are compared to solar abundances in Figure \ref{wdd2_gxp}.  Compared to the W7 model (model II), the production of Cr, Mn, Fe, Ni, Cu, and Zn isotopes seems to match the solar distributions more closely across the isotopic chains studied, though the underproduction of Cu and Zn still exists.

\begin{figure}[!]
	\includegraphics[width=\linewidth]{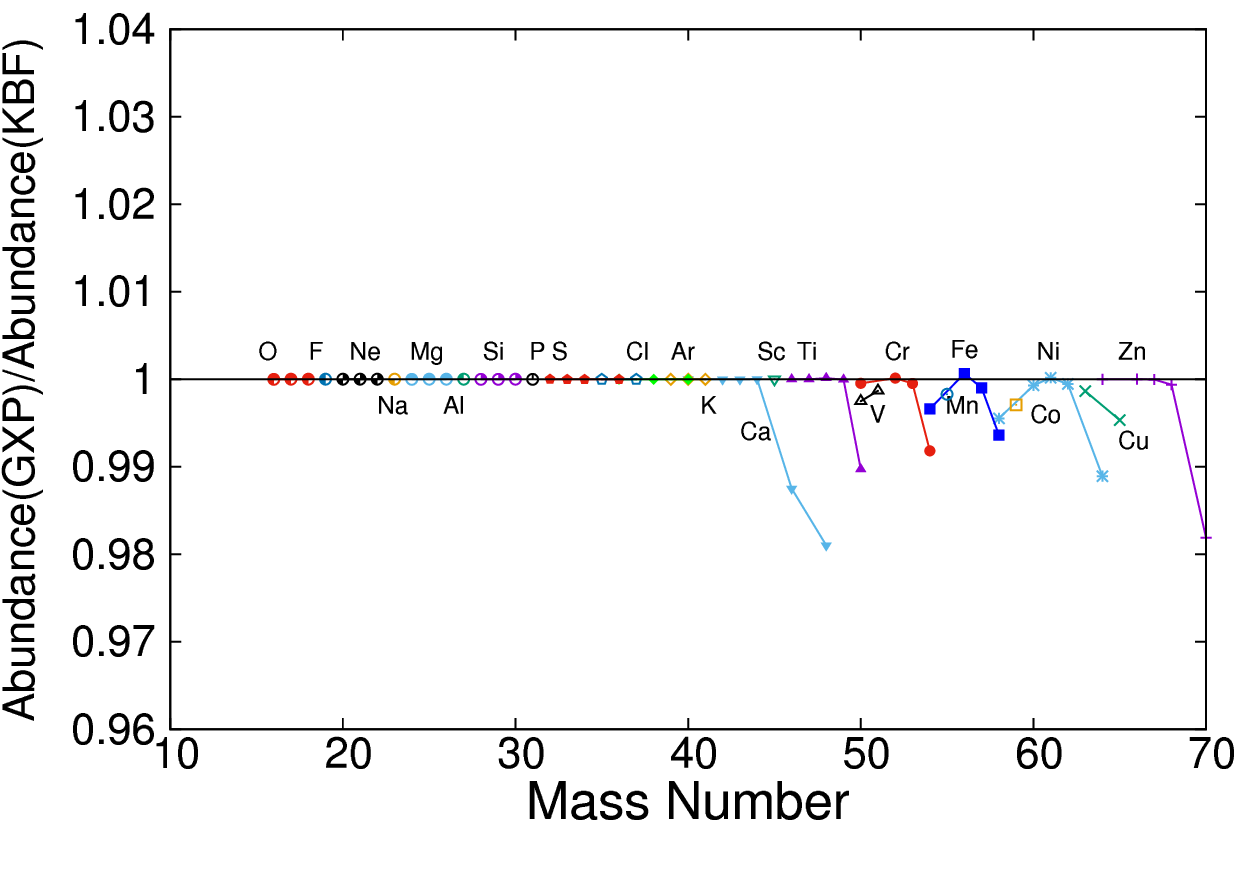}
	\caption{\label{wdd2_gxp_kbf}The ratio of abundances for nuclei produced in the GXP model to those of nuclei produced in the KBF model.  The results shown here are for the  
	WDD2 delayed-detonation model, comparing models III and IV in Table \ref{network_list}.}
\end{figure}

As with the deflagration model, the GXP calculation is compared to a calculation done using KBF rates for the delayed-detonation model.  The final abundance ratios are shown in Figure \ref{wdd2_gxp_kbf}.  
The ratios are slightly closer to unity in this model than in
	the deflagration model,  
	while the
		shapes are similar. This is an indicator that the delayed-detonation model is not as sensitive
		to rate changes as the deflagration model.  
The explanation for these small differences between the GXP and KBF rates is the same as that given in the comparison of the deflagration model.  
	That is, transitions 
in SNe Ia are only to low excitation energies as the Fermi energy of the electrons is $\sim$ 1 to 2 MeV. In this region, the differences between the GT strength functions are small.

A comparison of the total produced Fe and Ni mass is shown in Table \ref{mass_yields} for all four models studied.  One sees a slight shift to 
more $^{56}$Fe production using the GXP shell model, corresponding to a very slight increase in $Y_e$ as indicated in Table
\ref{fin_ye}.
\begin{table}
	\caption{\label{mass_yields}The total mass (in $M_\odot$) of $^{56}$Fe and $^{56}$Ni produced for each shell model and hydrodynamic model studied.
	Numbers in parentheses indicate the network-explosion model in Table \ref{network_list}.}
	\begin{tabular}{|l|c|c|c|c|}
		\hline
		~&\multicolumn{2}{c|}{W7}&\multicolumn{2}{c|}{WDD2}\\
		\cline{2-5}
		~&GXP(II) & KBF(I) & GXP(IV) & KBF(III) \\
		\hline
		$^{56}$Fe&0.66861&0.66631&0.68335&0.68291\\
		\hline
		$^{56}$Ni&0.65142&0.64911&0.66812&0.66764\\
		\hline
	\end{tabular}
\end{table}
\subsection{Nuclei in Excited States}
Given the temperatures of the environments of SNe Ia, it is conceivable that 
nuclei in their excited states may result in reaction rates that vary from 
the ground-state rates.  For this reason, it is worth mentioning the contribution to the EC rates for a few cases of nuclei in their excited
states.  

In Figure \ref{gtni56a_strength}, the GT strength functions are shown
in the GXP shell model for the first excited states of $^{56}$Ni and $^{54}$Fe, and
the integrated strengths are shown in Figure \ref{sumgt}.  In both cases, the first excited state of the parent nucleus is the 2$^+$ state. 
In each case, the strength function shifts to higher excitation energy.  The shift of the strength from
the excited 2$^+$ states can be understood as a 
manifestation of the Brink hypothesis, which is primarily applicable to giant dipole 
resonances and can be applied to GT strength distributions.  This hypothesis states that the 
giant dipole resonances (as well as the GT strengths) reside at the same energies
relative to excited states as in the ground state \citep{axel62,misch14,guttormsen16}.  
While not exact, it can be used to understand
the shift in the GT strengths in Figure \ref{gtni56a_strength}. Details concerning
the applicability of the Brink hypothesis have been 
discussed in \cite{martinez00}. Detailed 
studies of the applicability of the Brink hypothesis for
M1 strength functions are presented by \cite{loens12}, \cite{dzhioev14}, and \cite{dzhioev10}.
Thus, the shift of the strength
function is commensurate with the excitation energy of the parent nucleus.
In the case of SNe Ia nucleosynthesis, this shift would likely not change the resultant nucleosynthesis. This is because the excited states of the parent nuclei, the 2$^+$ state in both
$^{56}$Ni and $^{54}$Fe, are at 2.7 MeV and 1.4 MeV respectively.  At the temperatures
of the Type Ia SN environment, these states are not expected to have a significant population, and can be dismissed as possible contributors to the strength distributions.
\section{Discussion}
\label{discussion}
We continue the current effort in modern nuclear astrophysics towards a more accurate description of SN Ia nucleosynthesis using GT strength functions which more closely match experimental results.

Recent measurements of the GT strength distributions - used in 
determining EC rates - of
pf-shell nuclei have called into question the validity of the
theoretical distributions, which are important in determining
nucleosynthesis in SNe Ia \citep{sasano11,sasano12}. 

A comparison was made between the GT strength functions B(GT) 
of the commonly used KBF model \citep{kbf} and the GXP models
\citep{suzuki11}.  It was found that the 
GXP shell model more closely  matches the experimental B(GT) for $^{56}$Ni \citep{sasano11}.  

Using these strength distributions to determine EC rates for the pf-shell nuclei with 23$\le{Z}\le$30, 
nucleosynthesis in a type Ia SN was calculated using the thermodynamic trajectories 
of the deflagration model W7 \citep{nomoto84} and the delayed-detonation model 
WDD2 \citep{nomoto}. Comparisons of production were made between the W7 and WDD2 
models for rates computed using the GXP and KBF shell model parametrizations.

While the GXP shell model results in rates which tend to shift the resultant 
nucleosynthesis to higher $Y_e$, thus increasing
the Ni/Fe ratio, this shift is small, changing yields in the pf-shell region by 
only a few percent at best.
Even though B(GT) is dramatically different in each shell model calculation, resulting 
in potentially significant rate changes as shown in Figure \ref{ec_rate}, the overall 
change in rates is also a function of the electron Fermi distribution, which depends on the environmental temperature.
By the time the nucleosynthesis reaches the pf-shell nuclei, the temperature is low 
enough that the 
electron Fermi energy is about 1.5 MeV. Integrating over B(GT) in Figure \ref{gtni56a_strength} 
(while also taking the EC Q-value into account) results in a very small difference between
the resultant rates since the integration is only over small
transition energies, and the values of $E_x$ reached in the
GT strength distributions are small, where the deviation between shell models is insignificant.

Excited states for $^{56}$Ni and $^{54}$Fe have also been investigated, with a 
corresponding shift in the GT strength distribution.  Nuclei in excited states may shift the
EC Q-values, and the shift in the GT strengths (a manifestation of the Brink-Axel hypothesis) 
may result in a 
slight reduction of the EC rates.  However, it is noted that the excited states of 
doubly-magic $^{56}$Ni and singly-magic $^{54}$Fe in this regime are high enough that there will likely be little population of
these states in a SN Ia.  This should be studied in greater detail, however, as odd-A or 
odd-odd nuclei may have lower excited states with more significant populations in SNe Ia.

It's also worth noting that because the GXP 
	model is developed for pf-shell nuclei, 
application of a shell model with interactions which give different GT strengths to nuclei of lower mass, 
where the 
temperature of the SN Ia environment is hotter, may result in more significant changes in the nucleosynthesis.  
While type II 
supernovae are primarily responsible for production of these elements, SNe Ia are responsible for some of their 
production as well \citep{nomoto}.   
If the significant differences between the KBF and GXP models are at higher energies as suggested by the 
studies of $^{56}$Ni and $^{54}$Fe, then
low-mass nuclei may exhibit a more pronounced change in
abundance as they are produced earlier in the nucleosynthesis
when the temperatures are hotter.

Finally, the importance of $^{54}$Cr as a signature of the nucleosynthesis within the innermost 
mass shell is mentioned. This nucleus, with a Z/A=0.44, is one of the more proton-rich nuclei 
produced in the network calculations.  As such, it is only produced in any significant abundance
in the mass shells that have the lowest $Y_e$ distributions.  However, if the $Y_e$ distribution
is shifted globally over the entire star, then the production of $^{54}$Cr
may be altered dramatically. Any change to the central mass element (with the lowest $Y_e$) could
have relatively large effects on the global production of $^{54}$Cr.  It may be worthwhile to examine
this nucleus in greater detail and its sensitivity to the nuclear physics and stellar physics inputs.

In addition to $^{54}$Cr, it may be worthwhile to investigate $^{57}$Co and $^{55}$Fe as signatures of the Chandra 
and sub-Chandra models.  Recently, astronomers have detected radioactivity from $^{57}$Co and $^{55}$Fe \citep{shappee16,graur16}.
If these can be shown to be sensitive to the central density, then verifications from astronomical observations may help to constrain
the actual model used.

It is noted that corrections to EC rates for pf-shell nuclei relevant to SNe Ia using different nuclear shell models have resulted
in smaller relative effects in the overall nucleosynthesis
with historical changes decreasing with newer nuclear models.  
Because uncertainties in the nuclear physics are now apparently small, reducing the uncertainties in the stellar physics may provide larger corrections in the
nucleosynthesis computations.  Perhaps foremost among these would include an investigation of SNe Ia explosions
in three dimensions.  Also a systematic evaluation of the sensitivity of nuclear production to 
the uncertainties in the explosion may be worth pursuing.  One example of such an exploration is
the recent models in which the WD rotation is taken into account
\cite{benvenuto15}.  In these models the central density is predicted to have a large
variation and thus synthesis of neutron-rich Fe peak elements may be strongly affected.  This also suggests the general importance of accurate
electron capture rates for various conditions. Of course, uncertainties in nuclear physics relevant to other astrophysical scenarios,
	such as neutron-star crusts and type II supernovae, remain significant.
\acknowledgments
	KM's work was supported by NSF grant PHY-1430152;
	MAF's by NSF grant PHY-1204486 and by
	an NAOJ Visiting Research Professorship; 
	TK's by Grants-in-Aid for Scientific Research of JSPS 
	(26105517 and 24340060), of the Ministry of Education, Culture, Sports,
	Science and Technology of Japan; KN's by the World Premier International Research Center Initiative
	(WPI), MEXT, Japan, and by the Grants-in-Aid for Scientific Research
	of the JSPS (26400222 and 16H02168); and
	TS's was supported in part by Grants-in-Aid for Scientific Research (c)15K05090 of the MEXT of Japan.

\end{document}